\documentclass[a4paper]{jpconf}
\usepackage{iopams} \usepackage{aas_macros}
\usepackage{graphicx}
\begin{document}
\title{Genetic algorithms in astronomy and astrophysics}
\author{Vinesh Rajpaul$^{1,2}$}
\address{$^1$ Astrophysics, Cosmology and Gravity Centre (ACGC), Department of Astronomy, University of Cape Town, Private Bag X3, Rondebosch 7701, South Africa}
\address{$^2$ South African Astronomical Observatory, PO Box 9, Observatory, 7935, South Africa}
\ead{vinesh.rajpaul@uct.ac.za}
\begin{abstract}
Genetic algorithms (GAs) emulate the process of biological evolution, in a computational setting, in order to generate good solutions to difficult search and optimisation problems. GA-based optimisers tend to be extremely robust and versatile compared to most traditional techniques used to solve optimisation problems. This paper provides a very brief introduction to GAs and outlines their utility in astronomy and astrophysics.
\end{abstract}
\section{Introduction}
Many interesting mathematical problems can be reformulated as global optimisation problems; the solution of systems of algebraic or even differential equations, for example, can be cast quite naturally in terms of optimisation. The same holds true for the all-important inverse problems that are ubiquitous in the physical sciences, i.e.\ problems where one seeks to transform experimental data into model parameters in order infer properties of the physical systems being studied (a simple example: choosing parameters to minimise a $\chi^2$-statistic when fitting a Voigt-profile to a spectral line). 

The goal of a global optimisation problem is, given a so-called \emph{cost function}\footnote{Depending on the context, cost functions are also referred to as \emph{energy functions} or \emph{objective functions}.}  $f:\Omega \subseteq {\mathbb{R}^n} \to \mathbb{R}$, to try to find a point (more generally, a set of points) $\vec x^*\in\Omega$ such that:
\begin{equation}
\forall \vec x \in \Omega:f(\vec x) \geqslant f({ \vec x^*});
\end{equation}
$f(\vec x^*)$ is called the global minimum\footnote{Maximisation of $f(\vec x)$ is, of course, equivalent to minimisation of $g(\vec x):=-f(\vec x)$.}. A local minimum, $f(\hat {\vec x})$, is defined by the condition:
\begin{equation}
\forall \vec x \in \Omega,\exists \delta  > 0:\left\| {\vec x - \hat{\vec x}} \right\| < \delta  \Rightarrow f(\vec x) \geqslant f(\hat{\vec x}).
\end{equation}

Whereas finding an arbitrary local minimum of a function is a relatively straightforward task, especially if one has a good ``first guess'' -- extremely efficient techniques exist to solve such local optimisation problems -- finding \emph{global} minima is a far more challenging problem. Real-world cost functions tend to be nonlinear, discontinuous and/or hugely multimodal, and there is no foolproof approach to locating their global minima.

Most established approaches -- whether deterministic, stochastic or (meta)heuristic -- to solving global optimisation problems yield excellent results on a limited class of problems, but have drawbacks that tend to cripple them when faced with certain (reasonably) difficult problems. For example, they might get stuck too easily in local minima, they might be thwarted by discontinuous functions or they might be too slow to be of practical value when faced with enormous search spaces \cite{Charbonneau:2002b}. 

Evolutionary algorithms, inspired by biological evolution, are metaheuristic optimisation algorithms that tend to yield ``good enough'' results on a very wide range of (even extremely difficult) optimisation problems. So-called \emph{genetic algorithms} form one of the most successful subsets, and certainly the most popular subset, of evolutionary algorithms.
\section{Genetic algorithms: the basic idea}
\begin{figure*}[t]
	\begin{center}
	\includegraphics[scale=0.8]{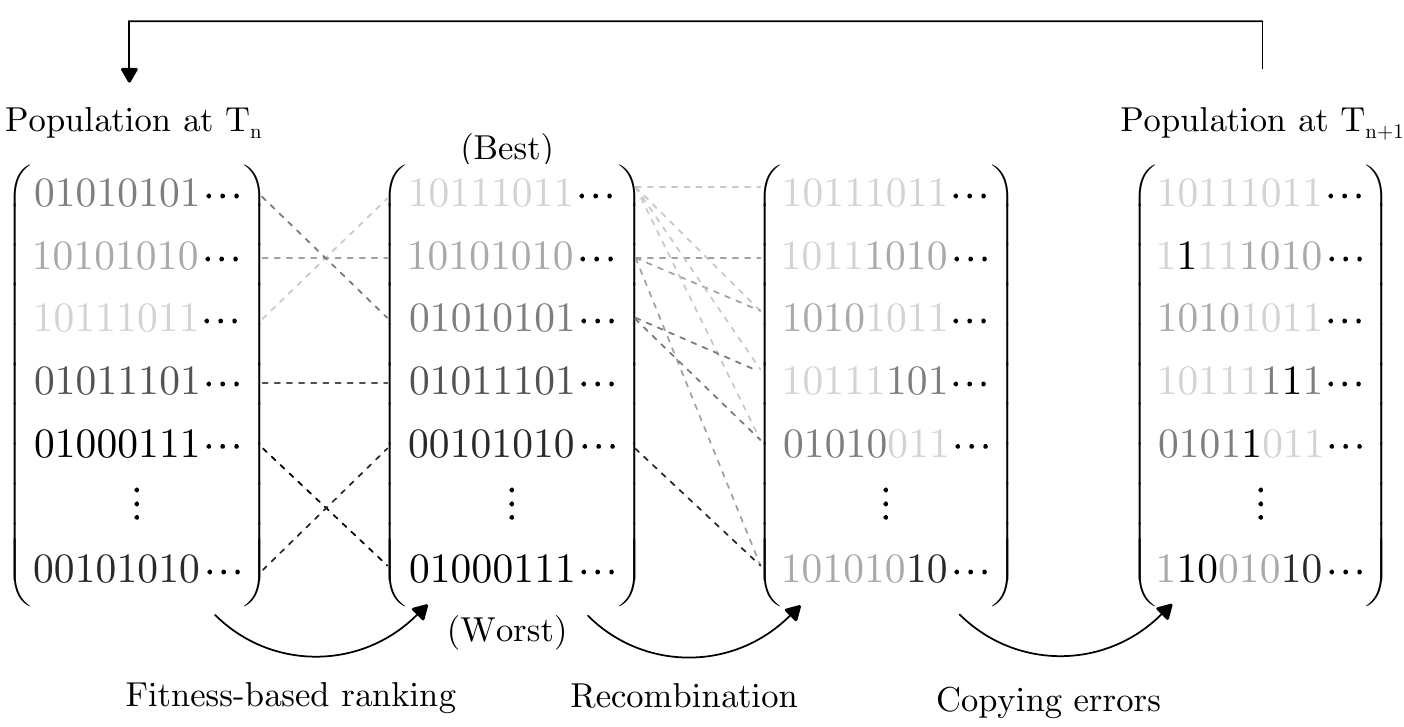}
	\caption{Schematic to illustrate the workings of a simple binary-coded genetic algorithm. Each bit represents a gene; here the genes of high-fitness solutions are given lighter colours.}
	\label{fig:schematic}
	\end{center}
\end{figure*}
Genetic algorithms, or GAs for short, draw inspiration from population genetics (and, like all evolutionary algorithms, from evolutionary biology in general) and they incorporate, in a computational setting, notions such as natural selection/survival of the fittest, genetic recombination, inheritance and mutation. The first GA-based optimiser was proposed in the mid-1970s \cite{Holland:1975}, and since then many modifications and improvements to the basic algorithm have been developed, including mechanisms without any direct biological analogues \cite{Haupt:2004}.

In spite of the rich variety of their potential incarnations, most GAs share a basic working scheme: they start with a population of many candidate solutions (called individuals or phenotypes), associate with each solution an encoded version of the phenotype (called a chromosome, genotype or an individual's genetic material) and also a measure of the solution's fitness (quality). This fitness function is often simply the additive inverse of the cost function to be optimised. Then, by repeated application of ``genetic operators'' mainly at the genotypic level, they cause the population as a whole to increase in phenotypic fitness, i.e.\ they cause the solutions to evolve towards optimality.

A typical (though simplistic and by no means general or optimal) working scheme for a genetic algorithm is as follows:
\begin{enumerate}
	\item[1.] construct a random initial population of genotypes;
	\item[2.] decode the genotypes and evaluate their phenotypic fitness; if the fittest phenotype matches the user-defined target fitness (or other termination criterion), \texttt{break}, otherwise continue;
	\item[3.] produce offspring by randomly selecting and recombining genetic material from the current population, favouring individuals with high phenotypic fitness;
	\item[4.] introduce, with some low probability, random changes (copying errors) into the genetic material of the offspring;
	\item[5.] replace low-fitness members of the old population with the offspring created in the previous step, and \texttt{goto} step 2.
\end{enumerate}
The selective recombination of the population's genetic material exploits good solutions to build even better ones, and the random mutations serve to inject entirely new and potentially favourable material into the gene pool that could not be obtained simply by recombining the genetic material of existing individuals.

Fig.\ \ref{fig:schematic} illustrates the working scheme of a simple GA where the solutions are encoded as binary strings\footnote{Most early GAs encoded solutions as binary strings, both for the sake of simplicity and supposed theoretical optimality; a large body of empirical evidence, however, indicates that it is preferable to work directly with floating-point representations of solutions when solving numerical optimisation problems \cite{Charbonneau:2002b,Haupt:2004,Michalewicz:1996,Wright:1991}.}. It may be shown that given enough time, and subject to a few reasonable assumptions, a GA will always converge to the global optimum of a cost function\footnote{Of course this knowledge is of little practical value; of more importance is the \emph{rate} of convergence to the global optimum, though unfortunately with GAs this rate is highly problem-dependent and difficult to estimate \emph{a priori}.} \cite{Michalewicz:1996,Eiben:1990}.
\section{GAs: pros and cons}
Relative to more conventional optimisation algorithms, GA-based optimisers offer a number of striking advantages, some of which are outlined below.

\emph{Robustness}. GA-based optimisers can handle -- with aplomb -- problems with multimodal
or low-contrast objective functions, multiple objectives and/or problems where the parameter
spaces have a very high dimensionality \cite{Charbonneau:2002b}.

\emph{Simplicity}. In order to solve a given optimisation problem, most ``off-the-shelf'' GAs requires only a single, unambiguous measure of the quality (fitness) of candidate solutions. They do not require, for example, gradients or Hessian matrices, the computation of which might be prohibitively difficult or impossible in some problems. Moreover the ideas underpinning GAs
are intuitively accessible and it is a relatively easy task to develop a working GA from scratch.

\emph{Speed}. Apart from the intrinsically high speed with which GAs tend to explore large parameter spaces \cite{Michalewicz:1996}, they are embarrassingly parallel: very little effort is required to transform a serial GA-implementation to a parallel implementation. Thus they are well-suited to exploiting high-performance hardware (multi-core workstations, graphics processing units, clusters etc.).

\emph{Versatility}. A single GA-based optimiser can be expected to yield ``good enough'' results on a very wide class of problems -- from a problem as simple as fitting a three-parameter Gaussian to some data, to one as complex as choosing a molecular configuration to minimise a Buckingham potential with hundreds of parameters -- and it is easy to incorporate problem-specific knowledge into a GA-based solver. The widespread adoption of GAs in fields such as engineering, chemistry, biology and economics bears testimony to their great versatility \cite{Haupt:2004}.
 
\begin{figure}[t]
\begin{minipage}[t]{0.43\linewidth}
\includegraphics[scale=0.8]{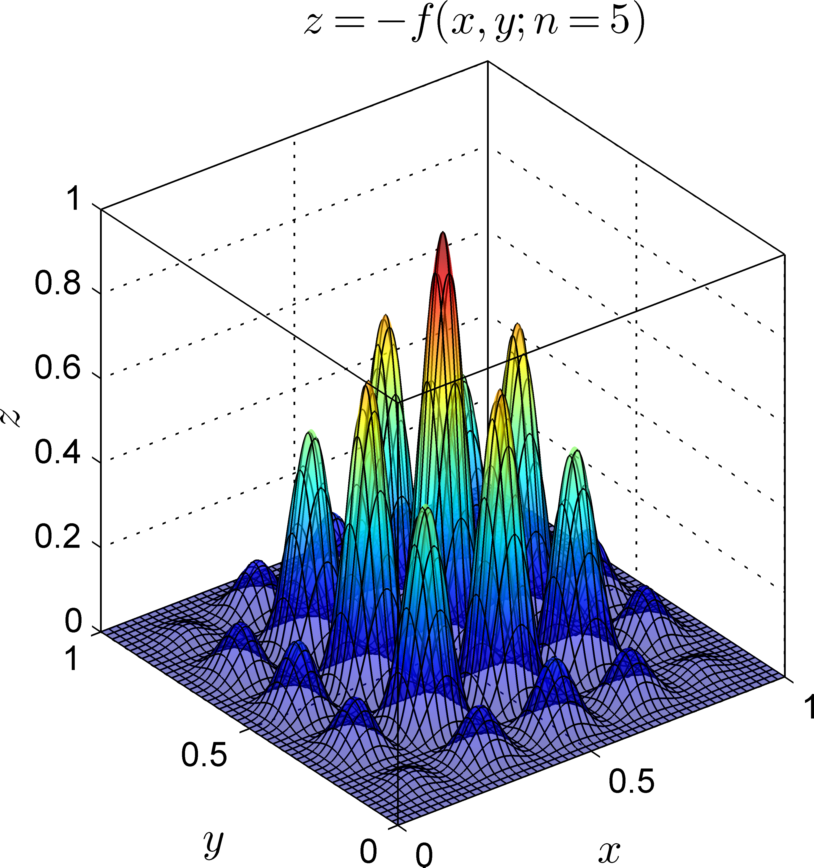}
\caption{Surface plot of the function $-f(x,y;n)$ defined by equation \ref{eq:Charbonneau}, for the case $n=5$. In this case there are $n^2=25$ local optima on the domain \hbox{$x,y\in[0,1]$}.}
\label{fig:landscape}
\end{minipage}
\hspace{0.2in}
\begin{minipage}[t]{0.51\linewidth}
\includegraphics[scale=0.8]{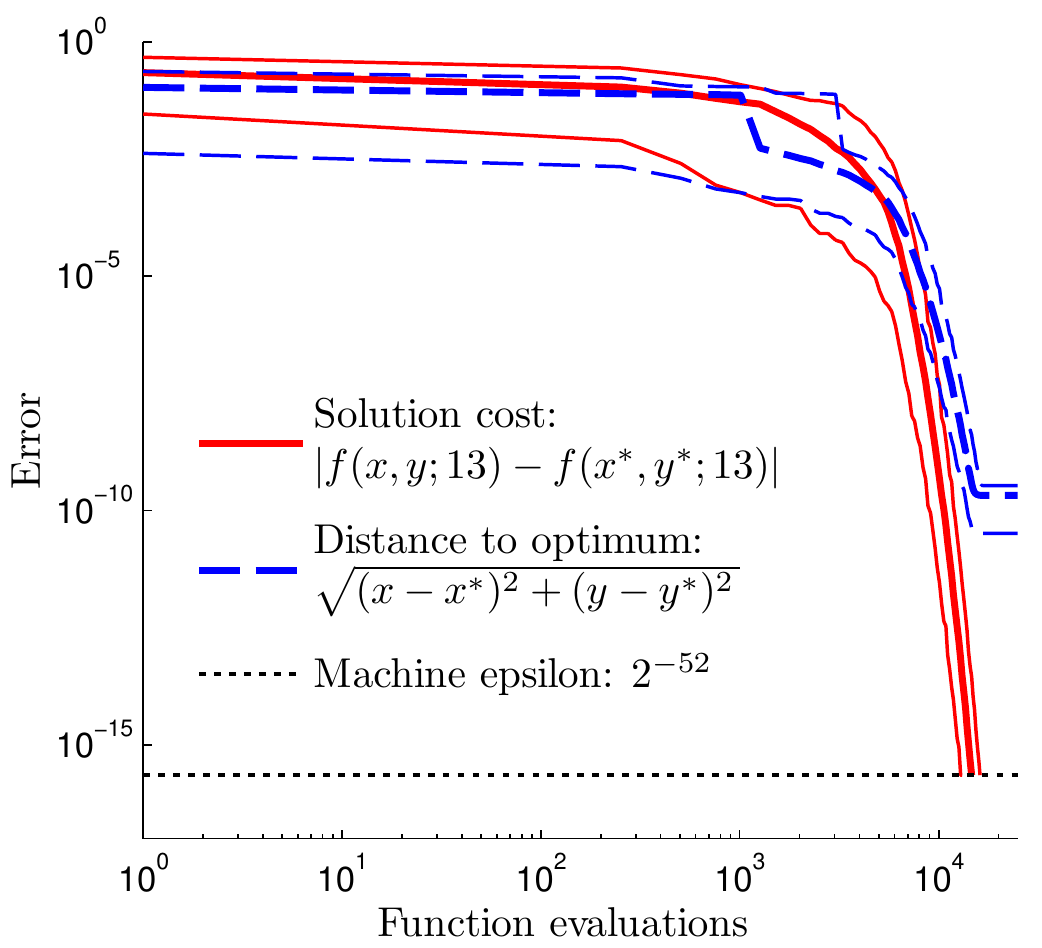}
\caption{Performance of a GA-based optimiser applied to the $n=13$ case of $f(x,y;n)$; the thick lines denote median performance in $10000$ trials, and the thin lines, upper and lower $3\sigma$ limits.}
\label{fig:performance}
\end{minipage}
\end{figure}

To illustrate the great robustness and versatility of a typical GA, consider the following cost function proposed by Charbonneau \cite{Charbonneau:1995}:
\begin{equation}
\label{eq:Charbonneau}
f(x,y;n) =  - {\left[ {16x(1 - x)y(1 - y)\sin (n\pi x)\sin (n\pi y)} \right]^2},
\end{equation}
where $x,y \in [0,1]$ and $n \in \{ 2k + 1;\forall k \in {\mathbb{N}}\}$. For $n=13$, say, it may be shown that $f(x,y)$ has $169$ (in general $n^2$) local minima on its domain, only one of which is the global minimum; moreover, the minima are separated by steep walls and there is little contrast between many of the minima (see fig.\ \ref{fig:landscape}).

Fig.\ \ref{fig:performance} illustrates how a GA-based optimiser fared on the (rather challenging) $n=13$ problem: in 10000 trials, the algorithm converged to the global minimum every single time, with the minimum location determined to a median accuracy of about one part in a billion after only $\sim10^4$ function evaluations (or a fraction of a second on a modern workstation). For comparison, a blind random search would require $\sim10^{18}$ evaluations to guarantee similar accuracy!

Although this performance is impressive in its own right, it is worth emphasising that it took mere minutes to adapt an existing GA-based optimiser\footnote{This GA, coded by the author, used floating-point encoding, dynamically-adjusted mutation rates and tournament-style selection of reproducing partners.} to solve this problem, and that the algorithm control parameters were not optimised in any way for this new problem.

An obvious question arises: \emph{why} do GAs work as well as they do? This topic is far beyond the scope of this survey paper but suffice it to say that a universally-accepted explanation has not yet been developed. Holland's famous \emph{Schema Theorem} has long been touted as providing an explanation for GAs' success \cite{Holland:1975}, although more recently it has become apparent that this theorem provides insight only into the workings of simplistic GAs; and even then, it is not clear whether the assumptions underlying the theorem are tenable \cite{Syswerda:1989,Wright:2003}.

Despite all their attractive features, GAs also have their share of disadvantages (more or less in accordance with Wolpert and Macready's famous ``no free lunch'' theorem \cite{NFL}). GAs might be called ``Jacks of all problems, but masters of none'': optimising a GA's performance on a given problem is often difficult or impossible, and in order to achieve near-optimal performance it is usually necessary to hybridise a GA with problem-specific heuristics. For example, they tend to be better at locating than at fine-tuning solutions: once a GA is in the vicinity of a global optimum, it is usually a good idea to let a local optimiser take over \cite{Charbonneau:2002b}. GAs can be inefficient on simple problems where the computational expense of applying the genetic operators outweighs that of evaluating the function to be optimised; conversely, on problems where each cost function evaluation is extremely expensive -- for example, where each evaluation requires a long simulation to be run -- a GA-based (or indeed any) forward modelling approach could be impractical. 

Finally the (currently) limited theoretical understanding of GAs is regarded by some, quite understandably, as a drawback and this might explain their relatively slow uptake in the physical sciences \cite{Charbonneau:1995}.
\section{Applications: astronomy and astrophysics}
This section presents a sample of the numerous and diverse applications that genetic algorithms have found in astronomy and astrophysics. For brevity's sake, only one or two short but representative examples have been drawn from different subfields.

\emph{Astrophysical dynamics}. Wahde and Donner developed a method for reliably determining the orbital parameters of interacting galaxies and applied their method to both artificial and real data \cite{Wahde:2001}. Their method is based on a GA that searches very efficiently through the large space of possible orbits; indeed, the authors argue that GAs are ideally suited for investigations of tidally interacting galaxies, where large multimodal spaces must be searched in order to constrain a large number of model parameters. Cant\'o \textit{et al}.\ devised an interesting variant of the canonical GA which they applied successfully to various problems, including the challenging task of finding the orbital parameters of the planets orbiting 55~Cancri, based on radial velocity measurements of the aforesaid stellar system \cite{Canto:2009}.

\emph{Physical and observational cosmology}. Although Monte Carlo methods seem to predominate in cosmology, GAs have already found a number of applications in the field. To mention just a few: Nesseris and Shafieloo used GAs to reconstruct the expansion history of the universe in a model-independent manner and thence, in conjunction with the so-called \emph{Om statistic}, they derived a null test on the cosmological constant model $\Lambda$CDM \cite{Nesseris:2010}; via GAs, Allanach \textit{et al}.\ were able to answer some important questions related to the discrimination of SUSY-breaking models, and in particular to quantify the measurements necessary to tell different SUSY-breaking scenarios apart \cite{Allanach:2004}; and Bogdanos and Nesseris used GAs to analyse Type \textrm{I}a SNe data and to extract model-independent constraints on the evolution of the dark energy equation of state \cite{Bogdanos:2009}. The latter authors note that as a non-parametric method, GAs provide a convenient model-independent platform for cosmological data analysis that can minimise bias due to premature choice of e.g.\ a dark energy model.

\emph{Gravitational lens modelling}. Gravitational microlensing is an ideal technique for probing the  galactic population of faint or dark objects such as substellar objects, stellar remnants and exoplanets. Though very successful, theoretical microlensing models tend to be complex and their associated inverse modelling problems are notoriously difficult. The author of this paper has recently been developing GAs to speed up this difficult modelling, with a view to being able to model ongoing events approximately in real time (i.e.\ on a timescale of minutes rather than weeks or months!) and thereby to facilitate better-informed observations and thus more useful observational data. Results of this work are expected to be published in early 2012. As another example, Liesenborgs \textit{et al}.\ presented a GA-based, non-parametric technique for inferring the projected lensing-mass distributions in strongly lensed systems \cite{Liesenborgs:2006}.

\emph{Stellar spectrum fitting}. Performing fits to stellar spectra is a nontrivial but important undertaking; from fitted models one can infer a veritable multitude of stellar properties. Baier \textit{et al}.\ were able to combine radiative transfer codes with a GA to produce an automated procedure for fitting the dust spectra of AGB stars. Their GA-based routine dramatically improved extant fits made with more traditional methods and provided a quantitative platform from which to compare different models \cite{Baier:2010}. In a similar vein, Mokiem \textit{et al}.\ used a parallelised GA as the basis for an autonomous fitter of spectra of massive stars with stellar winds \cite{Mokiem:2005}.

\emph{Stellar structure modelling}. Metcalfe and Charbonneau implemented a highly-parallelised and distributed GA to determine the globally optimal parameters for stellar models. The efficient, parallel exploration of parameter space made possible by their GA-based optimisation led to some important results in the field of white dwarf astroseismology, including the unexpected resolution of a then-puzzling discrepancy between stellar evolution model and astroseismic inferences of He-layer masses in DBV white dwarfs \cite{Metcalfe:2000}.

\emph{Telescope scheduling}. Autonomous telescope scheduling is a difficult task that requires dynamic adjustment of numerous observational constraints whilst trying to ensure the efficient achievement of many different scientific objectives. Kubanek developed an easy-to-implement yet robust approach to a robotic telescope scheduling problem, based on a GA that seeks out Pareto-optimal solutions (telescope schedules) \cite{Kubanek:2010}.
\section{Conclusions}
This paper introduced genetic algorithms, mentioned some of their strengths (and weaknesses) and finally illustrated their utility in astronomy and astrophysics. For those who would like to learn more about GAs or other evolutionary algorithms, there are many fine books and papers on the subject: to mention just a few, Michalewicz's book \cite{Michalewicz:1996} gives an excellent introduction with a theoretical leaning, and Haupt's book provides an equally good though more ``hands-on'' treatment \cite{Haupt:2004}. Goldberg's seminal book \cite{Goldberg:1989}, one of the most widely-cited works in all of computer science, serves as an outstanding tutorial-style reference, and finally Charbonneau's note \cite{Charbonneau:2002b} provides a straightforward discussion of how standard statistical methods can be used to construct confidence intervals for GA-estimated model parameters.

The author of this paper would welcome correspondence from anyone who would like to discuss evolutionary algorithms, perhaps with a view to applying them in their own work.
\ack
The author is grateful to the University of Cape Town and the National Research Foundation for the provision of financial support, and also to the anonymous referees for their helpful comments.
\section*{References}
\bibliographystyle{iopart-num}
\bibliography{SAIP}
\end{document}